\documentclass[twocolumn,pre,showpacs,amsmath,amssymb,reprint,superscriptaddress, nofootinbib]
{revtex4-1}

\usepackage{float}
\usepackage{graphicx}
\usepackage{amsmath,amssymb}
\usepackage{array}
\usepackage{color}
\usepackage{multirow}
\usepackage{scalerel}
\usepackage{bm}

\setcitestyle{round}
\usepackage[dvipsnames]{xcolor}
\usepackage{comment}

\usepackage{cleveref}

\usepackage{lineno}
\newcommand{\nPerp}{{\partial^{\perp}}}

\newcommand{\rl}{{r/\lambda}}

\newcommand{\etam}{{\eta_{2D}}}
\newcommand{\etaf}{{\eta_{3D}}}

\usepackage{xcolor}

\makeatletter
\newcommand{\globalcolor}[1]{%
  \color{#1}\global\let\default@color\current@color
}
\makeatother

\begin{document}
\def\hrSys{\scalerel*{\includegraphics{HpR.pdf}}{S}}
\def\rSys{\scalerel*{\includegraphics{R.pdf}}{S}}
\def\hSys{\scalerel*{\includegraphics{H.pdf}}{S}}
%\pagecolor{black}

\preprint{}

%\title{Hurricane dynamics in a membrane}
\title{Hyperuniformity and phase enrichment in vortex and rotor assemblies}
%from chaos to order, hidden order
%\title{Hurricane dynamics in a biomembrane}



\author{Naomi Oppenheimer}
\email{naomiop@gmail.com}
\affiliation{School of Physics, and the Center for Physics and Chemistry of Living Systems, Tel Aviv University, Tel Aviv 6997801, Israel}
\author{David B. Stein}
\affiliation{Center for Computational Biology, Flatiron Institute, New York, NY 10010, USA}
\author{Matan Yah Ben Zion}
\affiliation{School of Physics, and the Center for Physics and Chemistry of Living Systems, Tel Aviv University, Tel Aviv 6997801, Israel}
\affiliation{Laboratoire Gulliver, UMR CNRS 7083, ESPCI Paris, PSL Research University, 75005 Paris, France}
%\author{Haim Diamant??}
%\affiliation{Tel Aviv University}
\author{Michael J. Shelley}
\email{mshelley@flatironinstitute.org}
\affiliation{Center for Computational Biology, Flatiron Institute, New York, NY 10010, USA}
\affiliation{Courant Institute, New York University,
New York, NY 10012, USA}

\date{\today}

\begin{abstract}

Ensembles of particles rotating in a two-dimensional fluid can exhibit chaotic dynamics yet develop signatures of hidden order. Such ``rotors" are found in the natural world spanning vastly disparate length scales --- from the rotor proteins in cellular membranes to models of atmospheric dynamics. Here we show that an initially random distribution of either ideal vortices in an inviscid fluid, or driven rotors in a viscous membrane, spontaneously self assembles. Despite arising from drastically different physics, these systems share a Hamiltonian structure that sets geometrical conservation laws resulting in distinct structural states. We find that the rotationally invariant interactions isotropically suppress long wavelength fluctuations — a hallmark of a disordered hyperuniform material. With increasing area fraction, the system orders into a hexagonal lattice. In mixtures of two co-rotating populations, the stronger population will gain order from the other and both will become phase enriched. Finally, we show that classical 2D point vortex systems arise as exact limits of the experimentally accessible microscopic membrane rotors, yielding a new system through which to study topological defects.

\end{abstract}

%\keywords{}

\maketitle
%=================================================================================

\begin{figure*}[t]
\centering
\vspace{0.2cm} 
\includegraphics[width=0.8\textwidth]{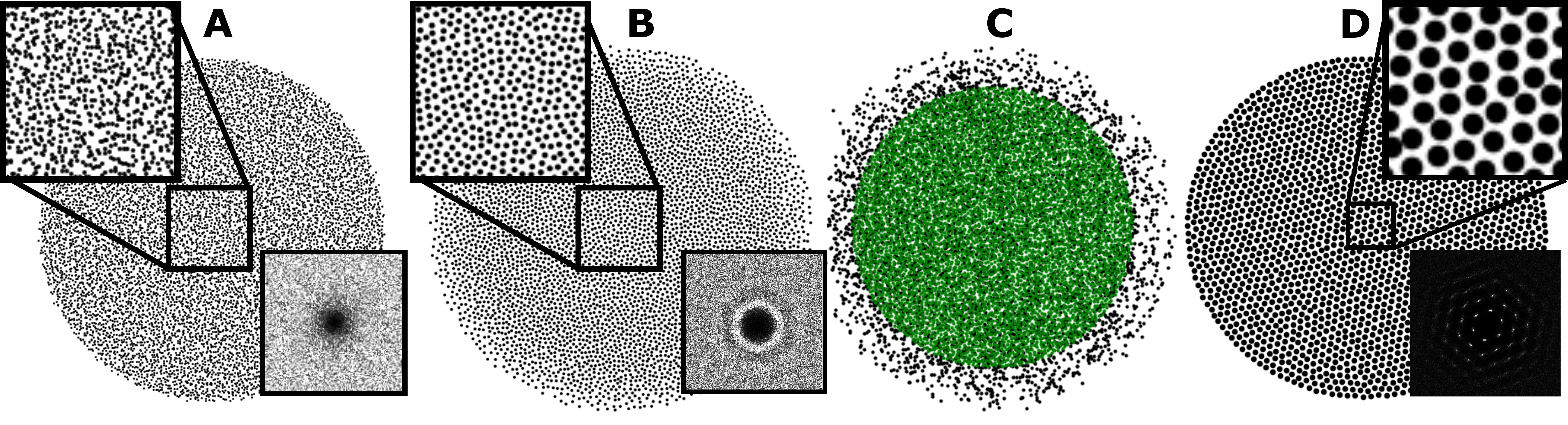} 
\caption{ Three different structural states of 2D vortices/rotors -
  hyperuniformity for Euler point vortices (A) and QG
  rotors/surface rotors (B), (C) phase enrichment induced by circulation differences where green (black) represents vortices of high (low) circulation, and (D)
  crystallization arising from hydrosteric interactions. The insets of (A), (B) and (C) show the structure
  factor, $S(q)$. In (A) and (B) $S(q)$ decays to zero
  at small $q$, indicating that the distribution is hyperuniform. In (C) the
  structure factor shows the six distinct peaks of a hexagonal
  lattice.}
\label{figPhases}
\end{figure*}
%switch the order and write first on conservation laws?
%Joanny, Onsager

Two-dimensional (or nearly so) fluid flows show rich and complex vortical dynamics. These can arise from flow
interactions with boundaries \cite{King1977,Shelley2011}, the inverse cascades of 2D turbulence \cite{Fjortoft1953, Kraichnan1967, Bernard2006}, from Coriolis force dominated atmospheric flows \cite{Behringer1991}, and from quantization effects in super
fluid He-II \cite{Abrikosov1957, Matthews1999}. Point vortices have long been staples for the modeling of such inertially dominated inviscid flows. Kirchoff \cite{Kirchhoff1876} was the first to describe point vortices using a Hamiltonian framework and his work was extended
by many others [e.g. \cite{Aref1983, Lin1941, Newton2001}],
notably, Onsager \cite{Onsager1949} in his statistical mechanics treatment
of 2D turbulence as clouds of point vortices.

Remarkably, structurally identical Hamiltonian and moment constraints can arise in the microscopic viscously-dominated realm from a strict
balance of dissipation with drive on immersed rotating objects. These
objects include models of interacting transmembrane ATP-synthase
``rotor-proteins'' \cite{Lenz2003,Lenz2004,Oppenheimer2019}, and the planar
interactions of rotors --- microscopic particles driven to rotate by an external torque \cite{Whitesides2000, Soni2019}.
%particles rotlets (a fundamental solution of the 3D Stokes
%equations) as models for microscopic particles driven to rotate
We refer to such systems as BDD
systems, as in balanced drive and dissipation. In modeling rotational BDD systems other physical effects may also come into play, such as steric interactions, that can yield interesting
complexities \cite{Oppenheimer2019}. Interacting assemblies
of driven-to-rotate particles has become an area of intensifying
interest in the active matter community
\cite{Whitesides2000,Lushi2014, Nguyen2014,Yeo2015,Goto2015,Soni2019, Bililign2021}.

Here we study both point vortices and a BDD rotor system of
rotationally-driven microscopic particles -- membrane rotors --
immersed in a flat membrane. We show that in both systems, their
Hamiltonian conservation laws lead to distinct structural states ---
hyperuniformity, phase enrichment and crystallization (see
Fig.~\ref{figPhases}), not yet observed for either system. We use the Hamiltonian to derive a bound for spatial correlations requiring hyperuniformity. We
demonstrate numerically that rotational dynamics robustly
self-assembles particles into a disordered hyperuniform 2D
material; This self-assembly is insensitive to the details of the hydrodynamic
interactions, steric repulsion, or the presence of impurities in the
form of different rotation rates. At steady state, the long wavelength configuration is characterized by an isotropically vanishing structure factor, $S({\bf q} \rightarrow 0) \rightarrow 0$ (where ${\bf q}$ is the wavevector), leading to an isotropic band-gap \cite{John1987,Yablonovitch1987, Man2013}. 

%These limits establish a precise connection between our microscopic BDD system and classical and central dissipation-free flow models.

%Rotating flows in 2D form rich
%patterns in an otherwise chaotic dynamics -- from the hexagon on
%Saturn's northern pole \cite{}, to quantization of vortices in super
%fluid He-II \cite{Donnelly1991}. Rotors are also found in everyday
%life --- the transmembranal ATP-synthase protein in the mitochondia of
%a biological cell
%\cite{Saffman1975,Singer1972,Lenz2003,Lenz2004,Buzhynsky2007,
%  Ueno2005}, hurricanes and quasi-geostrophical flows in the
%atmosphere \cite{Held1995,Falkovich2009, Cordoba2003}, and artificial
%magnetic particles rotated on a fluid interface \cite{Whitesides2000,
%  Lushi2014, Nguyen2014, Yeo2015,Goto2015,Soni2019}. Surprisingly the
%characteristics of membrane rotors are similar to those of point
%vortices. Surprising because membrane rotors and point vortices arise
%from vastly different physics -- membrane rotors move in the strict
%balance of dissipation and drive \cite{Saffman1975, Levine2004,
%  Seki2011, Camley2013, Oppenheimer2019}, while point vortices persist
%in the absence of any drive due to the complete lack of dissipation
%\cite{Newton2001, Saffman1975}. In the main text we focus on the duo
%---- membrane rotors/point vortices, but in the Supplementary
%Information (SI) we show that the results still apply to other 2D
%systems.

In classical mechanics, symmetries of the Hamiltonian
$\mathcal{H}$ restrict the phase-space of the conjugate variables,
position and momentum. However, in 2D point vortex or BDD point rotor
systems, the conjugate variables are the actual {\it spatial}
coordinates of the ensemble $\{x_i\}$ and $\{y_i\}$. The conservation
laws are therefore geometrical in nature, bounding the proximity and distribution of the particles. For both point vortices and
membrane rotors, as well as for a myriad of other 2D rotating systems
\cite{Whitesides2000, Nguyen2014, Lushi2014, Goto2015, Weijs2015,
  Soni2019}, the dynamics are dictated by Hamilton's equations,
\begin{equation}
\Gamma_i {\bf v_i} = \partial^{\perp}_i\mathcal{H},
\label{EqH}
\end{equation}
where $\partial^{\perp}_i = (\partial y_i, -\partial x_i)$, ${\bf v_i}$ is the velocity of rotor $i$, and
$\Gamma_i$ is the circulation (proportional to the magnitude of the torque for rotors).
Our finding, as we will show, is that the spatial arrangements of point vortices, as measured by $S({\bf q})$, are dictated by the Hamiltonian,
\begin{equation}
    \mathcal{H}[\rho({\bf r})] = \frac{N \Gamma^2}{4\pi} \int{\bf dq} \frac{S({\bf q})}{q^2}.
    \label{EqHSq}
\end{equation}

%A simple description of a hurricane is a point vortex in an ideal 2D fluid. 
To derive Eq.~\ref{EqHSq} and to find the Hamiltonian of $N$ particles, we first describe the flow due to a single vortex in an ideal Euler fluid and show its equivalence to a point rotor in a viscous membrane. We then use the linearity of the equations to extend the result to the many-body case. An ideal point vortex is given by a singular vorticity, ${\bf \omega} = \nabla \times
{\bf v} = \delta({\bf r})$. A 2D incompressible fluid can be described using a stream function $\Psi$ such that the velocity, ${\bf v}$, is given by ${\bf v} = \nPerp \Psi$. This equation, combined with the equation above gives,  $\Psi = - \frac{1}{2\pi} \log r$ \cite{Newton2001}. The flow, ${\bf v}(r)$, therefore, scales as $1/r$, where $r = |{\bf r}|$.

\begin{figure*}[tbh]
\vspace{0.2cm} 
\includegraphics[width=0.95\textwidth]{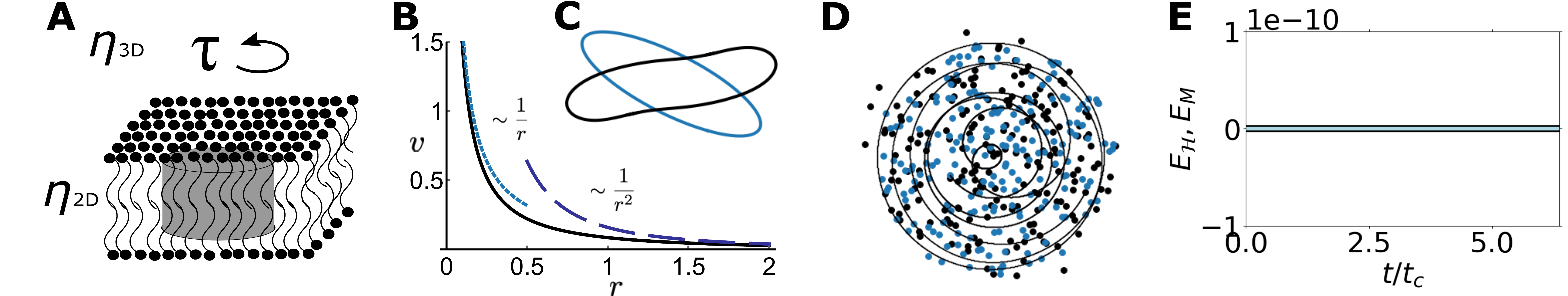} 
\caption{ 
%(a) a membrane rotor --- a disk of radius $a$ rotating with angular velocity $\Omega$ due to an external torque, and 
(A) A representation of a membrane rotor --- a disk rotating due to a torque $\tau$ in the plane of the membrane. (B) The velocity field due to a membrane rotor (solid line) which scales as a point vortex $v\sim 1/r$ at small distances (dotted), $\rl \ll 1$, transitioning to a QG behavior at large distances $v\sim 1/r^2$ (dashed). (C) Contour dynamics of an ellipse with radii ratios $r_l/r_s \leq 3$, where $r_l$ ($r_s$) is the major (minor) axis. Starting from the same contour, the dynamics differ according to the radius relative to the SD length. Blue is in the limit $r_l \ll \lambda$. In this limit the ellipse is rotating as a rigid body, as predicted by Kelvin \cite{Kelvin1880} for an elliptic patch in an Euler fluid. Black is in the limit $r_l \gg \lambda$, no longer conserving its shape since the large distance flow is in the quasigeostrophic regime. (D) 200 point membrane rotors, blue is the initial random configuration, black is the final configuration. Solid line shows typical trajectory of an individual vortex. Note that the area did not change considerably since the system of vortices is self-bounding. (E) the relative error in $\mathcal{H}$ and $M$ over a few cycle times, $t_c$. }
\label{figMembraneCartoon}
\end{figure*}

We switch now to a point rotor in a viscous membrane, driven by an external torque $\tau$. Following Saffman and Delbr\"uck's seminal work \cite{Saffman1975}, and many others that followed \cite{Levine2004, Seki2014, Camley2010}, we assume that the membrane is 
incompressible ($\nabla \cdot {\bf v}= 0$), and that inertia is negligible.  Under these assumptions, the Stokes momentum conservation equation for the membrane reads,
\begin{equation}
0 = \etam \nabla^2 {\bf v} + \etaf \left.\frac{\partial {\bf u^{\pm}}}{\partial z}\right|_{z=0^{\pm}} + \tau \nPerp \delta({\bf r}),
\label{eqMembranePoint}
\end{equation}
where ${\bf v}$ is the 2D velocity in the plane of the membrane,
${\bf u^{\pm}}$ is the 3D flow in the outer fluids, $\etam$ is the 2D viscosity, and $\etaf$ is the viscosity of the outer fluids. The second term on the
right hand side is the surface shear stress of the outer fluids, and the third term is
the force due to a rotating point object. There is no pressure
contribution when the motion is purely rotational. This equation is coupled to the equations of the outer fluids.
% $\nabla^2 {\bf u} = 0$, $\nabla
%\cdot {\bf u} = 0$ (note that there are no pressure gradients above an
%incompressible layer), and no-slip boundary conditions, ${\bf u}|_{z=0}
%= {\bf v}$. 
It is easy to solve the above equations using a 2D Fourier
Transform ($\widetilde{F}({\bf q}) = \int_{-\infty}^{\infty}
\int_{-\infty}^{\infty} F({\bf r}) e^{-i {\bf q} \cdot {\bf r}} d^2r$),
giving:
\begin{equation}
\widetilde{{\bf v}}({\bf q}) = \Gamma \nPerp \widetilde{\Psi} \ \ \ ; \ \ \  \widetilde{\Psi} = \frac{1}{q(q+\lambda^{-1})},
\label{EqStreamFourier}
\end{equation}
where $\Gamma = \tau/\etam$, and $\lambda = \etam/2\etaf$ is the Saffman Delbr\"uck length. At small distances ($r\ll \lambda$) momentum travels in the plane of the membrane. At large distances ($r\gg \lambda$) momentum travels through the outer fluid as well \cite{Oppenheimer2009, Oppenheimer2017}. 
%In real space $\Psi({\bf r}) = \pi/2 (H_0(\rl) - Y_0(\rl))$. 
In real space $\Psi({\bf r}) = 1/4 (H_0(\rl) - Y_0(\rl))$, where $H_0$ and $Y_0$ are zeroth order Struve function and Bessel function of the second kind respectively. 
In the limit of small distances, $r\ll \lambda$, the stream function is, $\Psi \approx -\frac{1}{2\pi} \log r$, i.e. {\it exactly} the same as for an ideal point vortex. In the opposite limit, $r\gg \lambda$, the stream function becomes $\Psi = \frac{1}{2\pi r}$ as in quasigeostrophic (QG) flows --- atmospheric or oceanic flows coming from gradients in pressure coupled to the Coriolis force \cite{Held1995}, or driven rotors on the surface of a fluid \cite{Yeo2015}. A membrane rotor, therefore, transitions from a point vortex for Euler at small distances to that of QG flow at large distances. The velocity is given by derivatives of $\Psi$ and is thus proportional to $1/r$ ($1/r^2$) in the limit of small (large) distances (see Fig.~\ref{figMembraneCartoon}B). 
For simplicity, we work primarily in the limit of small distances, $r \ll \lambda$, since in this limit the dynamics in a membrane converge with those of point vortices (many results still apply to the more general case). 
%as shown in the SI). 
In what follows, we will use ``point vortices" when there are only hydrodynamic interactions and ``rotors" when the particles have steric interactions in addition to hydrodynamic ones.

%\onecolumngrid
%\begin{center}
%\end{center}
%\twocolumngrid
The dynamics of $N$ point vortices follows from the Hamiltonian $\mathcal{H} = \frac{1}{2}\sum_{i\neq j} \Gamma_i \Gamma_j \Psi(|{\bf r}_i -{\bf r}_j|)$, where $\Gamma_i$ is the circulation of vortex $i$ (in a membrane $\Gamma_i = \tau_i/\etam$). The Hamiltonian depends on the conjugate variables ${\bf r}_i = (x_i,y_i)$, [normalized by the circulation $\sqrt{|\Gamma_i|}~{\rm sgn}(\Gamma_i) $], i.e. the positions of the vortices \cite{Newton2001}. The symmetries of the Hamiltonian correspond to conservation laws \cite{Noether1918}. In this case, we have symmetries with respect to translation in time, space, and rotation, corresponding to conservation of the Hamiltonian itself, and of the first and second moments of the distribution, ${\bf L} = \sum_i \Gamma_i {\bf r}_i (= {\bm 0}$ wlog), and $M = \sum_{i,j} \Gamma_i r_i^2$. Thus, the initial area cannot change dramatically, particles cannot drift to infinity since the second moment is fixed, nor can they collapse to a point since the Hamiltonian is conserved. These properties are readily observed in simulations. Figure~\ref{figMembraneCartoon}D shows typical trajectories of 200 membrane rotors. The initial distribution is random in a predefined finite area, and the dynamics are chaotic \cite{Aref1982}. The final configuration occupies nearly the same region of space as the initial configuration does, and the conservation laws hold to high precision in our simulations, as shown in Fig.~\ref{figMembraneCartoon}E. This self confining property of vortex dynamics has further consequences, as we now show.

\begin{figure*}[tbh]
\begin{center}
\vspace{0.2cm} 
\includegraphics[width=0.9\textwidth]{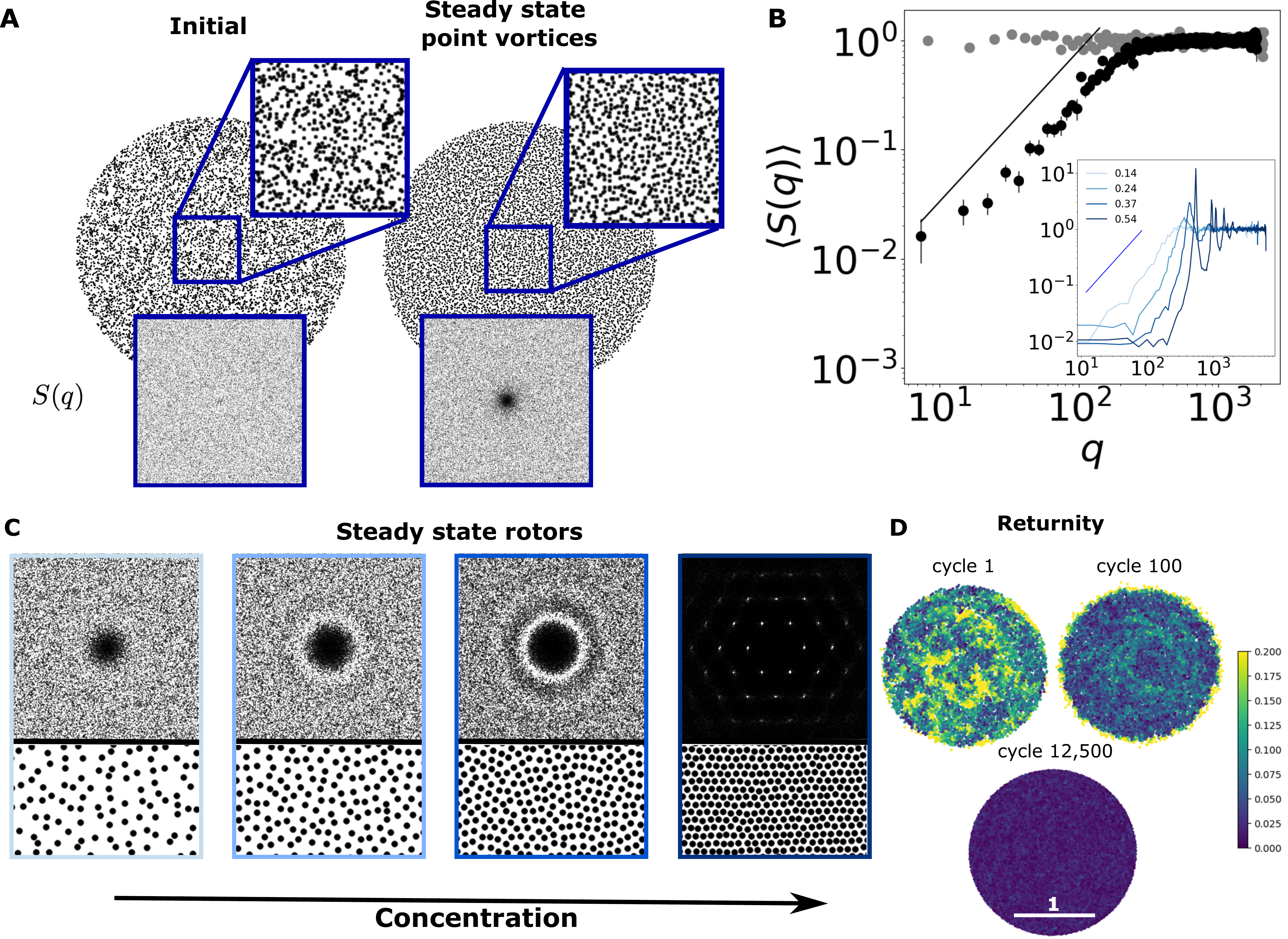} 
\caption{ 
%(a) a membrane rotor --- a disk of radius $a$ rotating with angular velocity $\Omega$ due to an external torque, and 
Hyperuniformity in ensembles of point vortices and rotors. (A) Snapshots of 10,000 point vortices initially (left) and at steady-state (right). Insets show the structure factor, $S({\bf q})$ showing a distinct cavity at steady-state. (B) Angular average of the structure factor shown in A, in a log-log scale with solid line showing a $q^{1.3}$ scaling. Error bars are standard deviation over 10 well separated timesteps. Inset shows the structure factor of the rotors shown in (C) with increasing hue corresponding to increased concentration $\phi = (0.14,0.24,0.37,0.54)$. Solid line is the same $\alpha \sim 1.3$ scaling. (C) Steady state configurations of 2,000 membrane rotors with the corresponding structure factors, showing a transition from disordered hyperuniformity to a hexagonal lattice. (D) A plot of the \textit{returnity} measuring the deviation of particle $i$ at position $r_i$ from its position at the previous cycle,  $returnity = \Delta r_i(t_{\rm cyc})/R$, where $R$ is the initial radius of the ensemble. The cycle time, $t_{\rm cyc}$, is defined at steady state as the distance between two adjacent minima of the function $f = \sum_i^N \Delta r_i(\Delta t)$, where $\Delta t$ is the time difference. Color scheme is from blue to yellow with increasing deviation.}
\label{figHyperuniformity}
\end{center}
\end{figure*}

\begin{center}
\begin{figure*}[tbh]
\vspace{0.2cm} 
\centering
\includegraphics[width=0.9\textwidth]{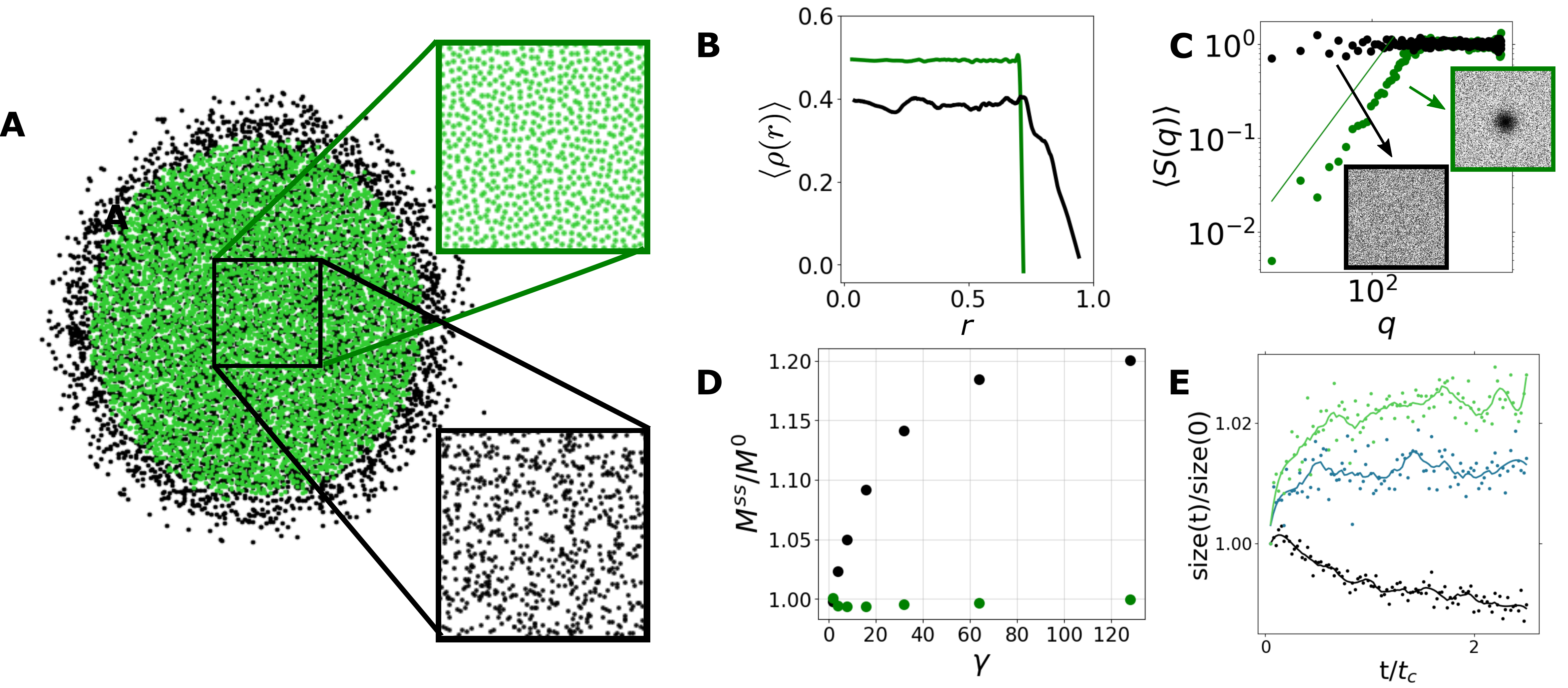} 
\caption{ 
Two populations of vortices with different circulations showing phase enrichment, $\Gamma_l=2\pi$ in black and $\Gamma_h=256 \pi$ in green. (A) Steady-state configuration for ten thousand point vortices of a circulation ratio $\gamma = \Gamma_h / \Gamma_l= 128$. Each inset shows a close-up view of one of the populations within the same physical region. (B) Density of the configuration in (A), $\rho(r)$, averaged over angle as a function of distance from the center. Note how density fluctuations are suppressed for the high circulation vortices, as is more clearly observed by the averaged structure factor, $S(q)$, in (C), where the solid green line shows a $\sim q^{1.3}$ power law. (D) The second moment for $N= 10,000$ vortices. Plotted separately for the high (in green) and low (in black) vortices at steady state as a function of $\gamma$ (i.e. increasing $\Gamma_h$). (E) LOSSLESS compression for the two populations showing an increase (decrease) in file size (an estimate of entropy) for the low (high) circulation vortices over a couple of cycles. In blue is the file size for the total system. Solid line is a moving average, time is normalized by an average cycle time $t_c$.}
\label{figSorting}
\end{figure*}
\end{center}

{\bf Hyperuniformity}. 
Hyperuniformity is the suppression of density-density fluctuations at small wavenumbers (or correspondingly, at large distances) \cite{Torquato2016, Hexner2015, Ariel2020}. Disordered hyperuniformity can emerge due to short ranged interactions such as those that arise in sheared suspensions \cite{Weijs2015,Wilken2020,Wang2018}, jammed materials \cite{Torquato2018}, and for spinning particles  \cite{Lei2019a}. Here we will show hyperuniformity emerging from long ranged interactions, similar to its emergence in sedimentation of irregular objects \cite{Goldfriend2017}. A good way to characterize hyperuniformity is the structure factor, defined as $S({\bf q}) =  N^{-1} |\widetilde{\rho}({\bf q})|^2$, where $\rho({\bf r}) = \sum_i \delta({\bf r} - {\bf r}_i)$ is the coarse grained density. In a hyperuniform material, $S(q)$ goes to zero as a power law at small wavevnumbers. % \xrightarrow{q\rightarrow 0} q^{\alpha}$ with a positive $\alpha$
We argue that point vortices must be hyperuniform due to the conservation of the Hamiltonian. For a density of rotors, the Hamiltonian is given by
$\mathcal{H}[\rho({\bf r})] \sim \frac{\Gamma^2}{2} \int {\bf dr} \int {\bf dr}'  \rho({\bf r}) \rho({\bf r}') \psi(|{\bf r} - {\bf r'}|).
$
%The number of rotors is also fixed, $\int_{\Omega}  \rho({\bf r}) {\bf dr} = 1$.
Using the convolution theorem, we find a general relation between the Hamiltonian and the structure factor
\begin{equation}
%    \mathcal{H}[\rho({\bf r})] = \frac{\Gamma^2}{4\pi}  \int{\bf dq} |\widetilde{\rho}({\bf q})|^2 \widetilde{\psi}({\bf q}) = \frac{N \Gamma^2}{4\pi} \int{\bf dq} S({\bf q})\widetilde{\Psi}({\bf q}).
    \mathcal{H}[\rho({\bf r})]  = \frac{N \Gamma^2}{4\pi} \int{\bf dq} S({\bf q})\widetilde{\Psi}({\bf q}).
    \label{EqHFourier}
\end{equation}
%This is a general result, which applies to the stream function of any Hamiltonian system. Specifically, 
In the case of point vortices, $\widetilde{\Psi}({\bf q}) = 1/q^2$, which gives Eq.~\ref{EqHSq}. For the integral of Eq.~\ref{EqHSq} to converge in 2D, $S({\bf q}) \sim q^{\alpha}$ near the origin, and we must have $\alpha>0$. In other words, an ensemble of point vortices is hyperuniform (a similar calculation in the QG limit, where $\widetilde{\Psi} = \lambda/q$, yields $\alpha > -1$).
Figures~\ref{figHyperuniformity}$B$ and \ref{figSorting}$C$, show an apparent $\alpha \sim {1.3}$ scaling for point vortices, consistent with the above argument. 

Using simulations we show that a set of $N$ vortices, uniformly distributed within a radius $R$, evolves to a disordered steady-state with a hidden order visible to the naked eye (compare Figures ~\ref{figHyperuniformity}${\rm A}$ left and right). We quantitatively characterize the system in steady-state in three ways: \textbf{(1)}~\textit{The structure factor.} 
At steady-state $S({\bf q})$ shows a distinct cavity, at $q\approx 0$, $S({\bf q}) \rightarrow 0$, for both points vortices (Fig.~\ref{figHyperuniformity}${\rm A}$) and rotors  (Fig.~\ref{figHyperuniformity}${\rm C}$). 
All simulations produce a hyperuniform arrangement. %However, for point vortices, it appears later and later as the time-step is decreased. It may be that perturbations are necessary to reach hyperuniformity, here very small but persistent time-stepping errors \cite{Dai1992}. 
\textbf{(2)}~\textit{Perturbations.} We demonstrate that  hyperuniformity is robust under different perturbations, be it in the form of numerical errors, repulsive interactions, or impurities (in the next section). For point vortices, the steady state appears later and later as the timestep is decreased, suggesting that perturbations are necessary for convergence, here very small but persistent timestepping errors \cite{Dai1992}. Adding steric interactions, hyperuniformity appears on a timescale that is independent of the timestep. 
Moreover, with steric interactions, as the area fraction $\phi$ of the particles is increased, the system transitions from disordered hyperuniform, to an ordered hyperuniform hexagonal lattice at $\phi \sim 0.5$, as can be seen in Fig.~\ref{figHyperuniformity}C. The inset of Fig.~\ref{figHyperuniformity}B shows the averaged structure factor where at intermediate area fractions we see Percus-Yevick type features for the structure factor of disks \cite{Percus1958}.  
 %We have simulated many configurations, changing step size and initial positions and hyperuniformity is always reached, though it seems a seed of randomness is necessary, as the time it takes to reach heyperuniformity is not linearly proportional to the circulation. 
\textbf{(3)}~\textit{The returnity.} We observe that at late times the ensemble of point vortices rotates almost as a rigid body and each particle goes back to its position at the previous cycle. We measure particle deviations by what we term the ``returnity" (see Fig.~\ref{figHyperuniformity}D for details). The system may seem to have reached an absorbing state, but the motion of vortices over many cycles is still chaotic. 

%\onecolumngrid
%\twocolumngrid

%\onecolumngrid
%\twocolumngrid
%Figure~\ref{figHyperuniformity}$B_2$ and \ref{figHyperuniformity}$C_2$ show a $q^2$ scaling for point vortices at the small $q$ limit. Hyperuniformity is therefore a steady state of ideal point-vortices and of rotors. The system reaches an absorbing state where the ensemble rotates almost as a rigid body, which we indicate by measuring what we term the "{\it Returnity}". The returnity $=(x_i(t=  ) ) $ Once steric repulsion is added, a faster decay is observed at intermediate wavelengths. We observe that collisions not only do not destroy hyperuniformity, they make it more pronounced and faster to reach as there is a combination of long-ranged and short-ranged interactions, thus adding stabilization to the system. Lastly, with steric interactions, as the concentration of particles is increased, the system transitions from disordered hyperuniform, to an ordered hexagonal lattice, as can be seen in Fig.~\ref{figHyperuniformity}D. 
%The absolute minimal entropical state must be a crystal. However, as we have shown before, a crystal is at most a neutrally stable state of the system~\cite{Oppenheimer2019}. In other words, it is not a state the system can reach spontaneously. Crystallization does occur but only when steric interactions are included, point particles will never spontaneously crystallize.

{\bf Rotation induced phase enrichment.}
We now show that for mixed populations of fast and slow rotating particles, there is phase enrichment of both populations and hyperuniformity of the fast ones. 
Consider a mixture of two equally numbered populations ($\rho_l = \rho_h$ at $t=0$) initially placed within the same radius $R$. $\rho_l$ rotates slowly with $\Gamma_l \ll \Gamma_h$, where $\Gamma_h$ is the circulation of the second population.
Figure~\ref{figSorting}A shows long-time simulation results for 10,000 point vortices. The two populations behave very differently. The fast vortices remain in a disk of only slightly smaller size than their initial area (Fig.~\ref{figSorting}B). The slow particle distribution shows a significant expansion. In addition, there is a striking difference when comparing the independently computed structure factors of these two populations, the fast vortices are hyperuniform with the same scaling as before, $S(q)\sim q^{1.3}$, whereas the slow ones show no signs of hyperuniformity (Fig.~\ref{figSorting}C). This difference is dramatic enough to be visible in a cursory examination of the separate distributions; see Fig.~\ref{figSorting}A.

Using a heuristic model, we show that the conservation laws allow two solutions at steady-state. In one solution, the two populations remain confined to a circle of the same radius. In the second solution, the radius of the slower population expands, while the radius of the faster population contracts. We then show that the segregated solution is the one that maximizes the number of states in the system. %We can analytically calculate the steady state radii assuming a homogeneous distribution of vortices. 
%Suppose there are two homogeneous densities of particles, initially both within a radius $R$. 
For simplicity, we assume that the final steady states are uniform (not true for the slow vortices as is clear from Fig.~\ref{figSorting}B).  There are two possible solutions where $\mathcal{H}$ and $M$ are conserved --- in the first, the initial radius, $R$, does not change; in the second, the radius of the fast vortices slightly decreases to $R_h$, allowing the slow vortices to expand to a larger radius $R_l$ given by $R_l^2 = (\gamma+1)R^2 - R_h^2\gamma$, where $\gamma=\Gamma_h/\Gamma_l$ (see Fig.~\ref{figSorting}D).
%One can see that even for a small compression of the fast rotors (i.e. $R_h \lesssim R$) y. 
Linearly expanding in $1/\gamma$, we find that $R_h \simeq R(1 - \beta/ \gamma)$ for the high circulation vortices, where $\beta$ is a positive prefactor of order 1. The slow vortices asymptote to $R_l \simeq R\sqrt{1+2\beta} + O(1/\gamma)$. The simulation results indicate that the outer radius indeed asymptotes to a larger valued constant as $\gamma$ increases and does not increase indefinitely (see Fig.~\ref{figSorting}D). %and SI).

A solution with two different radii is therefore possible and is indeed observed at large circulation ratios. Such a solution is favored entropically since it maximizes the available states. Asymptotically at large $\gamma$, the main entropical contribution is volumetric, $\Delta \mathcal{S}_{\rm volume} =  2 N \log(R_{\rm final}/R_{\rm initial})$. Since the high circulation vortices hardly change radius, $R_h \xrightarrow{\gamma\rightarrow \infty} R$, the change in entropy is coming mainly from the expansion of the low circulation vortices and is given by $\Delta \mathcal{S_{\rm total}} \sim N \log(1+2\beta) > 0$.
%By increasing their radius from $R$ to $R_l$, the slow particles increase their entropy by a considerable amount, while the fast vortices only slightly decreased theirs. 
Coupling the two populations allows one population to expand where before it was bounded \cite{OnsagerNote}. 
The situation is analogous to depletion interactions, where the net entropy of a system increases by condensing the large particles allowing for the small particles to explore a larger volume \cite{Kardar2007}.

%flows from the more negative one to the less so, in our case the high circulation vortices condense further, thus increasing their Hamilotnian, as a result they become even more negative and hyperuniform, the low circulation ones expand, thus lowering their free energy. 
%When temperature is negative, minimizing the free energy, Eq.~\ref{EqFreeEnergy}, means entropy must be minimized not maximized. 
%The absolute minimal entropical state in the infinite case is a crystal. However, a crystal is a neutrally stable state of rotors or of point vortices~\cite{Lenz2004, Oppenheimer2019}. In other words, it is not a state the system can reach spontaneously. However, crystallization can occur when steric interactions are included, as we see in Fig.~\ref{figHyperuniformity}F. In the absence of steric interactions the ensemble does not crystallize and instead entropy is decreased by the emergence of hyperuniformity. 
A simple way to estimate the entropy in a system is by using LOSSLESS compression, as suggested by Refs.~\cite{Martiniani2019, Avinery2019}. Compressing plots of particle positions in a system of 10,000 point vortices with circulation ratio $\Gamma_h/\Gamma_l=128$ shows an increase in file size for $\rho_l$ and a decrease for $\rho_h$, while the combined system is increasing, see Fig.~\ref{figSorting}E. 
  %Asymptotically, at large circulation ratio between the two populations (small $x$) the leading contribution The total entropy is $\Delta \mathcal{S} = \Delta \mathcal{S}_h + \Delta \mathcal{S}_l \approx 2N \log(1-\alpha x)+ N \log(1+ 2\alpha -\alpha^2 x) \approx N \log (1+2 \alpha - 2\alpha x - 6 \alpha^2 x)$, which is positive for $x < 1/(1+3\alpha)$. 
%The entropy of the fast rotors is slightly decreased due to two contributions: 1) They condense to $R_2 \leq R$, 2) since their density slightly increases they become hyperuniform.
%Out of all these contribution the change in area of the slow particles is the most significant. The situation is reminiscent of depletion interactions where small particles condense large ones in order to increase their free volume.

{\bf Discussion.}
We have shown that driven particles in a membrane or a soap film, as well as point vortices in an ideal 2D fluid, have geometrical conservation laws which limit their distribution. These conservation laws dictate different possible structural states --- namely hyperuniformity and phase enrichment. We have shown that hyperuniformity is robust to several forms of perturbations whether arising due to numerical errors, steric interactions, or impurities in the form of low circulation vortices.  For rotors with steric interactions, the unbounded ensemble crystallizes into a hexagonal lattice when the area fraction $\phi\gtrsim0.5$ (see also \cite{Oppenheimer2019}). We have limited the discussion to membrane rotors and vortices, but the results hold for other settings in which mass is conserved in the 2D plane, e.g. particles at the surface of a fluid.% (see SI).

What is especially interesting about our particular BDD system is its
potential for experimental realizability, its moment and Hamiltonian
structure, and that its near-field interactions (i.e. below the
Saffman-Delbruck length) are identical to those of Euler point
vortices. Further, the far-field interactions of membrane rotors are
identical to those of point vortices of the semi-quasigeostrophic
equations \cite{Held1995, Falkovich2009, Cordoba2003} used to model
atmospheric flows.  Thus, to observe the interesting dynamical
features we describe, one does not need to go to the atmospheric
scale, or cool a fluid to near-zero temperature. In principle, one can
simply observe microscopic particles on a soap film, in smectic films, a membrane, or even at the surface of a fluid
\cite{Nguyen2010, Lumay2013, Yeo2015, Soni2019}.  
%Finally, the Euler-Lagrange equation, combined with the equation for the vorticity $\nabla^2 \Psi = \omega$ gives,
%\begin{equation}
%    \nabla^2 \Psi = \Gamma e^{-\alpha -\beta \Gamma \Psi - \Gamma \gamma r^2}.
%\end{equation}

%Joyce and Montgomery used Onsager's argument to derive a mean field equation for the density of vorticity, which applies for rotors as well
%\begin{equation}
%    \nabla^2 \psi = \sum_i \Gamma_i \rho_{i0} e^{-\beta \Gamma_i \psi},
%\end{equation}
%where $\rho_{i0}$ is a normalization factor of the $i$th species such that $\int d^2r \rho_i = N_i/N$.

 {\bf Methods.}
 \begin{footnotesize}
 {\it Simulations.} Simulations were performed in Python. Random initial configurations within the unit disk were found by rejection sampling (points in the unit rectangle were sampled uniformly, transformed to the rectangle $[-1,1]^2$, and those with $r>1$ were discarded). The initial Hamiltonian $H_0$ is computed at $t=0$, and the relative error $\epsilon(t)=|H_t-H_0|\textbf{}/H_0$ is monitored as a measure of fidelity. For simulations of rotors (i.e. with steric repulsion), a 5th order explicit Runge-Kutta method based on the Dormand-Prince scheme  \cite{dormand1980family} with a fixed timestep size of $\delta t=10^{-7}$ was used. Long integration times were required for simulations of point vortices, and for these simulations an exlpicit eighth-order adaptive method based on the Dormand-Prince scheme \cite{prince1981high,solving_odes} was used, with both relative and absolute tolerances set to $10^{-6}$. The specific implementation of the scheme used was the \emph{DOP853} method of \emph{scipy.integrate} \cite{2020SciPy-NMeth}. For simulations of 10,000 point vortices with $\Gamma=2\pi$, $\epsilon(t)<1.6\times10^{-3}$ up to $t\approx 16,000 $ cycles, while for simulations with 5,000 vortices with $\Gamma=2\pi$ and 5,000 vortices with $\Gamma=256\pi$, $\epsilon(t)< 5\cdot 10^{-3}$ up to $t \approx 10$ cycles. Time is normalized by the average cycle time, $t_c \approx 4\pi^2 R^2/\sum_i\Gamma_i$, where $R$ is the initial radius. %For two populations $t_c$ was approximated numerically. 
 
%  The simulations were performed in Python using either: (i) 7/8th order Runge-Kutta with adaptive time stepping and a tolerance of $rtol = 1\cdot 10^{-6}$, for the hyperuniformity simulations of point vortices and for the phase enrichment simulations, or (ii) 4/5th order Runge-Kutta with a fixed time-step of $dt=10^{-7}$, for the hyperuniformity simulations of rotors and for some the phase enrichment simulations. Initial configurations are always random. The Hamiltonian for 10,000 point vortices is conserved up to a relative error of $(H_t - H_0)/H_0 = 1.5 \cdot 10^{-3}$ at physical time $t=10$ corresponding to a million time steps. For two populations with a circulation ration of $x=1/32$ with adaptive timestepping, the hamitlonian it is conserved up to a relative error of $10^{-3}$ at time $t=1.2$. For circulation ratio $x/128$, with a fixed time step it is conserved up to a relative error of $5\cdot 10^{-3}$. 
Steric interactions were taken as the repulsive part of a harmonic potential, i.e. for two particles whose centers are distance $r_{i}$ apart, $F = -k r_{ij}$ if $r_{ij}<2a$ and zero otherwise. The use of a harmonic potential, rather than a sharp step function for hard core particles, provided improved numerical stability and convergence. A large $k$ value was chosen to ensure no overlap between particles, $k = 1\cdot 10^6$, for particles of size $a = 0.01$.

 {\it Structure factor.} To accurately compute the structure factor $S({\bf q})$ we use a type-1 non-uniform fast-Fourier transform \cite{barnett2019parallel}. Explicitly, points are restricted to a windowing region which is confined entirely within the unit disk. The frequencies $\widetilde\rho({\bf q})$ are computed for the first 512 modes in each direction, and the average value (i.e. $\widetilde{\rho}(0)$) is set to $0$. This results in structure factors in the plane, such as those shown in Fig.~\ref{figHyperuniformity}. Except in those cases where crystallization occurs, these structure factors are azimuthally isotropic. To summarize this information, the angular average over the structure factor was calculated by slicing the result to 1000 equal bins between $q_{\rm min}$ and $q_{\rm max}$ and taking the mean of the results that fell within each slice.

 {\it Compression.} A plot of the positions of the point vortices was compressed using PNG with AGG backend. Each vortex was plotted by a single pixel. The total size of the plots was kept fixed in time. The figure size was chosen to minimize overlap between neighboring vortices but maintaining a computationally accessible file size.

\end{footnotesize}

{\bf Acknowledgment}
We thank Haim Diamant for insightful discussions regarding the emergence of hyperuniformity from the conservation laws, to Martin Lenz for suggesting a simple heuristic model of the phase enrichment, and to Enkeleida Lushi. N.O. acknowledges supported by the Israel Science Foundation (grant No. 1752/20). M.J.S. acknowledges support by the National Science Foundation under Awards Nos. DMR-1420073 (NYU MRSEC), DMS-1620331, and DMR-2004469.
\newpage
\bibliography{huricaneDynamics} 
\bibliographystyle{apsrev4-1}

\end{document}